\documentstyle[pra,aps]{revtex}
\def\mathbf{\vec}

\def\ca{\c{c}\~{a}}

\begin{document}
\draft
\title{'t Hooft Determinant:
       Fluctuations and Multiple Vacua}
\author{Brigitte Hiller and Alexander A. Osipov\thanks{On leave from the Joint Institute for
 Nuclear Research, Laboratory of Nuclear Problems, 141980 Dubna,
        Moscow Region, Russia.} }
\address{Centro de F\'{\i}sica Te\'{o}rica, Departamento de
         F\'{\i}sica da Universidade de Coimbra, 3004-516 Coimbra, Portugal}
\date{\today}
\maketitle






\begin{abstract}
The 't Hooft six quark flavor mixing interaction ($N_f=3$) is bosonized by 
the path integral method. The considered complete Lagrangian is constructed 
on the basis of the combined 't Hooft and $U(3) \times U(3)$ extended
chiral four fermion Nambu -- Jona-Lasinio interactions. The method of
the steepest descents is used to derive the effective mesonic Lagrangian.
Additionally to the known lowest order stationary phase (SP) result of 
Reinhardt and Alkofer we obtain the contribution from the small quantum 
fluctuations of bosonic configurations around their stationary phase 
trajectories. Fluctuations give rise to 
multivalued solutions of the gap equations, marked at instances by drastic 
changes in the quark condensates, a novel scenario for the vacuum state 
of hadrons at low energies. 
\end{abstract}


\section{Introduction}
The global $U_L(3)\times U_R(3)$ chiral symmetry of the QCD Lagrangian 
(for massless quarks) is broken by the $U_A(1)$ Adler-Bell-Jackiw 
anomaly of the $SU(3)$ singlet axial current $\bar{q}\gamma_\mu\gamma_5q$.
Through the study of instantons \cite{Hooft:1976,Diakonov:1995},
it became clear that effective $2N_f$ quark
interactions, known as 't Hooft interactions, violate $U_A(1)$, but are still invariant 
under chiral $SU(N_f)\times SU(N_f)$.In the
case of two flavors they are four-fermion interactions,
and the resulting low-energy theory resembles the old Vaks Larkin Nambu
Jona-Lasinio model \cite{Nambu:1961}. In the case of three flavors
they are six-fermion interactions which are responsible for the correct
description of $\eta$ and $\eta'$ physics, and additionally lead to the
OZI-violating effects \cite{Bernard:1988,Kunihiro:1988},
\begin{equation}
\label{Ldet}
    {\cal L}_{\mbox{det}}=\kappa (\mbox{det}\bar{q}P_Rq
                                 +\mbox{det}\bar{q}P_Lq)
\end{equation}
where the matrices $P_{R,L}=(1\pm\gamma_5)/2$ are projectors and the determinant
is over flavor indices.  

The physical degrees of freedom of QCD at low-energies are mesons. The
bosonization of the effective quark interaction (\ref{Ldet}) by the
path integral approach has been considered in \cite{Reinhardt:1988}, where 
the lowest order stationary phase approximation (SPA) has been used to 
estimate the leading contribution from the 't Hooft determinant.    
In this approximation the functional integral is dominated by the 
stationary trajectories $r_{\mbox{st}}(x)$, determined by the extremum 
condition $\delta S(r)=0$ of the action $S(r)$. The lowest order SPA  
corresponds to the case in which the integrals associated with 
$\delta^2 S(r)$, for the path $r_{\mbox{st}}(x)$ are neglected and
only $S(r_{\mbox{st}})$ contributes to the generating functional. 
The subject of the present work is to include the contribution associated with $\delta^2 S(r)$.
Our motivation is the following: let us consider the one-dimensional integrals of
the form
\begin{equation}
   F(\lambda )=\int^\infty_{-\infty}dt \exp [i\lambda f(t)]   
\end{equation}
What is sought by the method of SP is the dominant contribution to $F$ 
as $\lambda\rightarrow\infty$. The dominant contribution to the integral
comes from regions of $t$ where $f'$ vanishes. Supposing that $f'$
vanishes at only one point $t_0$ and neglecting contributions to the
integral from regions far from $t_0$, one can obtain the result
\begin{equation}
   F(\lambda )=\sqrt{\frac{2\pi i}{\lambda f''(t_0)}}\ e^{i\lambda
   f(t_0)}.
\end{equation}
Hence $F(\lambda )$ goes to zero like $1/\sqrt{\lambda}$. The lowest 
order SPA would correspond to the result $F(\lambda )=const\cdot\exp 
[i\lambda f(t_0)]$, which has the incorrect asymptotic behavior $F(\lambda )=
{\cal O}(1)$. If results related to finite-dimensional integrals, such as 
$F(\lambda )$, mean anything with regard to corresponding functional 
integrals, one can conclude that the leading SPA should include the 
contribution from the integrals associated with the second functional 
derivative $\delta^2 S(r_{\mbox{st}})$. 
\section{Path Integral Bosonization}
To be definite, let us consider the theory of the quark fields in four
dimensional Minkowski space, with dynamics defined by the Lagrangian
density
\begin{equation}
\label{totlag}
  {\cal L}={\cal L}_{\mbox{NJL}}+{\cal L}_{\mbox{det}}.
\end{equation}
The first term here is the extended version of the Nambu -- Jona-Lasinio 
(NJL) Lagrangian ${\cal L}_{\mbox{NJL}}={\cal L}_0+{\cal L}_{\mbox{int}}$, 
consisting of the free field part 
\begin{equation}
  {\cal L}_0=\bar{q}(i\gamma^\mu\partial_\mu -\hat{m})q,
\end{equation}
and the $U(3)_L\times U(3)_R$ chiral symmetric four-quark interaction  
\begin{equation}
  {\cal L}_{\mbox{int}}=\frac{G}{2}[(\bar{q}\lambda_aq)^2+
                        (\bar{q}i\gamma_5\lambda_aq)^2].
\end{equation}
We assume that quark fields have color and flavor indices running
through the set $i=1,2,3$; $\lambda_a$ are the standard $U(3)$ Gell-Mann 
matrices with $a=0,1,\ldots ,8$. The current quark mass, $\hat{m}$, is a 
nondegenerate diagonal matrix with elements $\mbox{diag}(\hat{m}_u, 
\hat{m}_d, \hat{m}_s)$, it explicitly breaks the global chiral 
$U(3)_L\times U(3)_R$ symmetry of the ${\cal L}_{\mbox{NJL}}$ Lagrangian. 
The second term in (\ref{totlag}) is given by (\ref{Ldet}). Letting  
\begin{equation}
\label{param}
   s_a=-\bar{q}\lambda_aq, \quad 
   p_a=\bar{q}i\gamma_5\lambda_aq, \quad 
   s=s_a\lambda_a, \quad
   p=p_a\lambda_a
\end{equation}
yields 
\begin{equation}
\label{mdet}
   {\cal L}_{\mbox{det}}=-\frac{\kappa}{64}\left[
   \mbox{det}(s+ip)+\mbox{det}(s-ip)\right].
\end{equation}

The dynamics of the system is described by the vacuum transition
amplitude in the form of the path integral
\begin{equation}
\label{genf1}
   Z=\int {\cal D}q{\cal D}\bar{q}\exp\left(i\int d^4x{\cal L}\right).
\end{equation}  
which can be written in the equivalent form \cite{Reinhardt:1988}
\begin{equation}
\label{genf3}
   Z=\int {\cal D}\sigma_a{\cal D}\phi_a
          {\cal D}q{\cal D}\bar{q}
          \exp\left(i\int d^4x{\cal L}_q\right)
     \int {\cal D}r_{1a}{\cal D}r_{2a}
          \exp\left(i\int d^4x{\cal L}_r\right)
\end{equation}  
where
\begin{eqnarray}
\label{lagr2}
  {\cal L}_q(\bar{q},q,\sigma ,\phi )
&=&\bar{q}(i\gamma^\mu\partial_\mu -\hat{m}-\sigma 
              +i\gamma_5\phi )q, \\
\label{lagr3}
  {\cal L}_r(\sigma ,\phi ,r_1,r_2)&=&2G\left[(r_{1a})^2+(r_{2a})^2\right]
              -2(r_{1a}\sigma_a+r_{2a}\phi_a) \nonumber \\
            &-&\frac{\kappa}{8}\left[
              \mbox{det}(r_{1}+ir_{2})
              +\mbox{det}(r_{1}-ir_{2})\right].
\end{eqnarray}
The Fermi fields enter the action bilinearly, we can always integrate
over them.  At this stage one should also shift the scalar fields 
$\sigma_a\rightarrow\sigma_a+\Delta_a$ by demanding that the vacuum 
expectation values of the shifted fields vanish $<0|\sigma_a|0>=0$,
yielding gap equations to fix parameters $\Delta_a = m_a -
\hat{m}_a$, where $m_a$ denotes the constituent quark masses 
\footnote{The shift by the current quark mass is needed to hit the 
correct vacuum state, see e.g. \cite{Osipov:2001}.}. 
To evaluate path integrals 
over $r_{1,2}$ one has to use the method of stationary phase, or,
after the formal analytic continuation in the time coordinate $x_4=ix_0$,
the method of steepest descents. The analytical continuation of the Euclidean  version of the path integral under consideration is  
(see \cite{Osipov:2002} for details),

\begin{equation}
\label{ancJ}
     J(\sigma ,\phi )=
     \int^{+i\infty+r_{\mbox{st}}}_{-i\infty +r_{\mbox{st}}}
     {\cal D}r_{1a}{\cal D}r_{2a}
     \exp\left(\int d^4x{\cal L}_r(\sigma ,\phi ,r_1,r_2)\right).
\end{equation} 
Near the saddle point $r^a_{\mbox{st}}$,
\begin{equation}
\label{lagr4}
  {\cal L}_r\approx {\cal L}_r(r_{\mbox{st}})
            +\frac{1}{2}\sum_{\alpha ,\beta }\tilde{r}_\alpha
            {\cal L}''_{\alpha\beta}(r_{\mbox{st}})\tilde{r}_\beta 
\end{equation}
where $r^a_{\mbox{st}}$ is a solution of the equations
${\cal L}'_r(r_1,r_2)=0$ determining a flat spot of the surface 
${\cal L}_r(r_1,r_2)$, 
\begin{equation}
\label{saddle}
  \left\{
         \begin{array}{rcl}
         2Gr^a_1-(\sigma +\Delta )_a
         -\frac{3\kappa}{8}A_{abc}(r_1^br_1^c-r_2^br_2^c)&=&0 \\
         2Gr^a_2-\phi_a+\frac{3\kappa}{4}A_{abc}r_1^br_2^c&=&0.
         \end{array}
  \right.
\end{equation}
with totally symmetric constants, $A_{abc}$, closely related with the $U(3)$ constants $d_{abc}$. We use 
in (\ref{lagr4}) symbols $\tilde{r}^a$ for the differences
$(r^a-r^a_{\mbox{st}})$. To deal with the multitude of integrals in
(\ref{ancJ}) we define a column vector $\tilde{r}$ with eighteen
components $\tilde{r}_\alpha =(\tilde{r}^a_1, \tilde{r}^a_2)$ and
with the matrix ${\cal L}''_{\alpha\beta}(r_{\mbox{st}})$ being equal to
\begin{equation}
\label{Qab}
  {\cal L}''_{\alpha\beta}(r_{\mbox{st}})=4GQ_{\alpha\beta},
  \quad
  Q_{\alpha\beta}=\left(
  \begin{array}{cc}
  \delta_{ab}-\frac{3\kappa}{8G}A_{abc}r_{1\mbox{st}}^{c}
  &\frac{3\kappa}{8G}A_{abc}r_{2\mbox{st}}^{c}\\
  \frac{3\kappa}{8G}A_{abc}r_{2\mbox{st}}^{c}
  &\delta_{ab}+\frac{3\kappa}{8G}A_{abc}r_{1\mbox{st}}^{c}
  \end{array}
  \right).
\end{equation} 
Eq.(\ref{ancJ}) can now be concisely written as
\begin{equation}
\label{ancJ2}
     J(\sigma ,\phi )=\exp\left(\int d^4x {\cal L}_r(r_{\mbox{st}})
                      \right)
     \int^{+i\infty}_{-i\infty}
     {\cal D}\tilde{r}_{\alpha}
     \exp\left(2G\int d^4x\tilde{r}^{\mbox{t}}Q(r_{\mbox{st}})
     \tilde{r} 
     \right)\left[1+{\cal O}(\hbar )\right].
\end{equation} 

The solutions of Eqs.(\ref{saddle}) are the following even and odd parity combinations 
$r^a_{1\mbox{st}}$ and $r^a_{2\mbox{st}}$ expressed in the form of increasing
powers in $\sigma_a , \phi_a$  
\begin{eqnarray}
\label{rst}
   r^a_{1\mbox{st}}&=&h_a+h_{ab}^{(1)}\sigma_b
       +h_{abc}^{(1)}\sigma_b\sigma_c
       +h_{abc}^{(2)}\phi_b\phi_c+\ldots \\
   r^a_{2\mbox{st}}&=&h_{ab}^{(2)}\phi_b
       +h_{abc}^{(3)}\phi_b\sigma_c
       +\ldots 
\end{eqnarray}
Putting these expansions in Eqs.(\ref{saddle}) one obtains a series 
of selfconsistent equations to determine the constants $h_a$,
$h^{(1)}_{ab}$, $h^{(2)}_{ab}$, etc. \cite{Osipov:2002}
As a result we get
\begin{equation}
\label{lam}
   {\cal L}_r(r_{\mbox{st}})=-2h_a\sigma_a
                             -h_{ab}^{(1)}\sigma_a\sigma_b  
                             -h_{ab}^{(2)}\phi_a\phi_b
                             +{\cal O}(\mbox{field}^3).
\end{equation}

We now turn to the evaluation of the path integral in Eq.(\ref{ancJ2}). 
In order to define the measure ${\cal D}\tilde{r}_\alpha$ more 
accurately we expand $\tilde{r}_\alpha$ in a Fourier series
\begin{equation}
   \tilde{r}_\alpha (x)=\sum^\infty_{n=1}c_{n,\alpha}\varphi_n(x),
\end{equation} 
assuming that suitable boundary conditions are imposed. The set of 
the real functions $\{\varphi_n(x)\}$ form an orthonormal and complete 
sequence, therefore 
\begin{equation}
\label{ancJ3}
     \int{\cal D}\tilde{r}_{\alpha}
     \exp\left(2G\int d^4x\tilde{r}^{\mbox{t}}Q(r_{\mbox{st}})
     \tilde{r}\right)=\frac{C}{\sqrt{\det (2G\lambda^{\alpha\beta}_{nm})}}.
\end{equation}
with
\begin{equation}
\label{Qab2}
   2G\lambda^{\alpha\beta}_{nm}
   =\left(
            \begin{array}{cc}
            h^{(1)-1}_{ac}&0\\
            0&h^{(2)-1}_{ac}
            \end{array}
    \right)_{\alpha\sigma}
            \left(\delta_{\sigma\beta}\delta_{nm}
            +\int d^4x\varphi_n(x)F_{\sigma\beta}(x)\varphi_m(x)
            \right)
\end{equation}
and
\begin{equation}        
            F_{\sigma\beta}=\frac{3\kappa}{4}A_{eba}\left(
                   \begin{array}{cc}
                   -h^{(1)}_{ce}(r^a_{1\mbox{st}}-h_a)&
                   h^{(1)}_{ce}r^a_{2\mbox{st}}\\
                   h^{(2)}_{ce}r^a_{2\mbox{st}}&
                   h^{(2)}_{ce}(r^a_{1\mbox{st}}-h_a)
                   \end{array}
                   \right)_{\sigma\beta}.  
\end{equation}
Only the matrix $F_{\sigma\beta}$ depends here on fields $\sigma , \phi$. 
By absorbing in $C$ the irrelevant field independent part of 
$2G\lambda^{\alpha\beta}_{nm}$, and expanding the logarithm in the
representation $\det (1+F)=\exp\mbox{tr}\ln (1+F)$, one can obtain 
finally for the complete action 
\begin{equation} 
\label{Sra} 
    S_r=\int d^4x\left\{   
      {\cal L}_r(r_{\mbox{st}})+
      \frac{a}{2G^2}\sum_{n=1}^\infty\frac{(-1)^n}{n}\mbox{tr}\left[
      F^n_{\alpha\beta}(r_{\mbox{st}})\right]\right\}
\end{equation}
proposing that the undetermined dimensionless constant $a$ will be fixed
by confronting the model with experiment afterwards \cite{Jackiw:2000}

\section{The Ground State}
Let us study the ground state of the model under consideration, then 
properties of the excitations will follow naturally.
In Eq.(\ref{lagr3}) the field coefficients $h_i$ obey
\begin{equation} 
\label{hi}
   2Gh_i-\Delta_i=\frac{\kappa}{8}t_{ijk}h_jh_k
\end{equation}
with the totally symmetric coefficients $t_{ijk}$ equal to zero
except for $t_{uds}=1$ and  with order parameters 
$\Delta_i\neq 0\quad (i=u,d,s)$.

A tadpole graphs calculation gives for the gap equations the following 
result
\begin{equation}
\label{gap}
   2h_i+\frac{3a\kappa}{8G^2}
        \left(h^{(2)}_{ab}-h^{(1)}_{ab}\right)A_{abc}
        h^{(1)}_{ci}=\frac{N_c}{2\pi^2}m_iJ_0(m_i^2)
\end{equation}
where the left hand side is the  contribution from (\ref{Sra}) and the right
hand side is the contribution of the quark loop from (\ref{genf3}), \cite{Osipov:2002}.



The second term on the left hand side of Eq.(\ref{gap}) is the 
correction resulting from the Gaussian integrals of the steepest
descent method, comprising the effects of small fluctuations around 
the stationary path. If one puts for a moment 
$a=0$ in Eq.(\ref{gap}), and combines the result with Eqs.(\ref{hi}), 
one finds gap equations which are very similar to the ones obtained in
\cite{Bernard:1988}. 
For this case one obtains values of $(m_u,m_s)$ which are uniquely related to values of $(G,\kappa )$.

The general case, which we have when $a>0$ in 
Eq.(\ref{gap}), yields in turn  a 
region for $m_u,m_s$, where three 
values of couplings $(G,\kappa)$ are possible. 

Conversely, one can study the solutions: $m_u=m_u(G,\kappa ),\ 
m_s=m_s(G,\kappa)$ at fixed values of input parameters: 
$\Lambda ,\hat{m}_u=\hat{m}_d,\hat{m}_s$. Again we find several extremal solutions. 
The minima will be identified in a future analysis of the effective potential.
We refer to \cite{Osipov:2002} for graphical displays.

Multivalued solutions of the gap equations have been obtained in a different context in \cite{Bicudo:2002}.

\section{Concluding remarks}
The purpose of this work has been twofold. Firstly we have developed
the technique which is necessary to go beyound the lowest order SPA
in the problem of the path integral bosonization of the 't
Hooft six quark interaction.  This technique is rather general and can be
readily used in other applications. Second, we have explored with
considerable detail the implications of taking the quantum
fluctuations in account in the description of the hadronic vacuum.  We
encountered several classes of solutions. These multiple vacua may have very interesting physical consequences and
applications.  
\vspace{0.5cm}

{\bf Acknowledgemnets:} This work is supported by Funda\ca o para a Ci\^encia e a Tecnologia, POCTI/35304/FIS/2000 and NATO "Outreach" Cooperation Program.
\vspace{0.5cm}

\baselineskip 12pt plus 2pt minus 2pt


\begin{thebibliography}{99}
\bibitem{Hooft:1976} A. M. Polyakov, Phys. Lett. B 59 (1975) 82;
        Nucl. Phys. B 121 (1977) 429. 
        A. A. Belavin, A. M. Polyakov, A. Schwartz and Y. Tyupkin,
        Phys. Lett. B 59 (1975) 85.
        G. 't Hooft, Phys. Rev. Lett. 37 (1976) 8; Phys. Rev. D 14
        (1976) 3432.
        C. Callan, R. Dashen and D. J. Gross, Phys. Lett. B 63 (1976)
        334.
        R. Jackiw and C. Rebbi, Phys. Rev. Lett. 37 (1976) 172.
        S. Coleman, ``The uses of instantons'' Erice Lectures, 1977. 
\bibitem{Diakonov:1995} D. Diakonov, ``Chiral symmetry breaking by 
        instantons'', Lectures at the Enrico Fermi School in
        Physics, Varenna, June 27 - July 7 (1995); {\tt hep-ph/9602375}.
\bibitem{Nambu:1961} V. G. Vaks, A. I. Larkin ZhETF 40 (1961) 282;
        Y. Nambu, G. Jona-Lasinio, Phys. Rev. 122 (1961) 345; 124 (1961) 246.
\bibitem{Bernard:1988} V. Bernard, R. L. Jaffe, U.-G. Mei\ss ner, Nucl. Phys.
        B 308 (1988) 753.
\bibitem{Kunihiro:1988} T. Kunihiro and T. Hatsuda, Phys. Lett. B 206
        (1988) 385.
        T. Hatsuda, Phys. Lett. B 213 (1988) 361.
        Y. Kohyama, K. Kubodera and M. Takizawa, Phys. Lett. B 208
        (1988) 165.
        M. Takizawa, Y. Kohyama and K. Kubodera, Prog. Theor. Phys. 82
        (1989) 481. 
\bibitem{Reinhardt:1988} H. Reinhardt and R. Alkofer, Phys. Lett. B 207
        (1988) 482.
\bibitem{Osipov:2001} A.A. Osipov and B. Hiller, Phys. Rev. D 62
        (2000) 114013; idem Phys. Rev. D 63 (2001) 094009.
\bibitem{Osipov:2002} A.A. Osipov and B. Hiller, Phys. Lett. B 539 (2002) 76.
\bibitem{Jackiw:2000} R. Jackiw, Int. J. Mod. Phys. B 14 (2000) 2011;
        {\tt hep-th/9903044}.
\bibitem{Bicudo:2002} P.J. Bicudo, J.E. Ribeiro and A.V. Nefediev, Phys. Rev. D 65
        (2002) 085026.
\end{thebibliography}
\end{document}